\documentclass[twocolumn,showpacs,amsmath,amssymb,aps,floatfix]{revtex4}

\usepackage{graphicx}
\usepackage{dcolumn}
\usepackage{bm}

\begin{document}
\title{First Passage Time Distribution of multi-scale stationary Markovian processes}        
\author{Salvatore Miccich\`e}
\affiliation{Dipartimento di Fisica e Tecnologie Relative, Universit\`a degli Studi di Palermo, \\
             Viale delle Scienze, Ed. 18, I-90128 Palermo, Italy}

\date{\today}

\begin{abstract} 

The aim of this paper is to investigate how the correlation properties of a stationary Markovian stochastic processes affect the First Passage Time distribution. First Passage Time issues are a classical topic in stochastic processes research. They also have relevant applications, for example, in many fields of finance such as the assessment of the default risk for firms' assets.  

By using some explicit examples, in this paper we will show that the tail of the First Passage Time distribution crucially depends on the correlation properties of the process and it is independent from its stationary distribution. When the process includes an infinite set of time-scales bounded from above, the FPTD shows tails modulated by some exponential decay. In the case when the process is power-law correlated the FPTD  shows power-law tails $1/t^{(\alpha+1)/2}$ and therefore the moments $\langle t^n \rangle$ of the FPTD are finite only when $n< (\alpha-1)/2$. We will also show that such power-law behaviour is not merely due to the fact that the process includes an infinite and unbounded set of time-scales. Rather, the time-scale must enter the FPTD with weights that must be distributed according to a power-law for large time-scales values. 

Finally,  we will give a general result connecting the FPTD of an additive stochastic processes $x(t)$ to the FPTD of a generic process $y(t)$ related by a coordinate transformation $y=f(x)$ to the first one.

\end{abstract}

\maketitle
%
%
%
\section{Introduction}   \label{intro}

Stochastic processes are used to model a great variety of systems in disciplines as disparate as physics \cite{BM, ornstein,VanKampen81,risken,gardiner,Schuss,Oksendal}, 
genomics \cite{Waterman,Durbin}, finance \cite{Bouchaud,Mantegna}, climatology \cite{vanStorch} and social sciences \cite{Helbing}. 

Here we will devote our attention to the special class of Markovian stochastic processes that are also stationary and that can be described by a Fokker-Planck (FP)
equation \cite{VanKampen81,risken,gardiner}. Such processes are fully determined by the knowledge of their probability density function (pdf)
$W(x,t)$  and their conditional transition probability $P(x_{n+1},t_{n+1}|x_{n},t_{n})$. 

The simplest stationary Markovian stochastic process is the Ornstein-Uhlenbeck (OU) one that is characterized by an exponential autocorrelation function $e^{-t/T}$ where $T$ is referred to as the time-scale of the  process. For a general stationary Markovian additive process the methodology of eigenfunction expansion \cite{risken,gardiner} allows us to write the autocovariance function as the infinite weighted sum of exponential functions $e^{-\lambda t}$ each characterized by a time-scale $\lambda^{-1}$, where the $\lambda$ values are the eigenvalues of the FP operator. By analogy with respect to the OU process, for an additive stationary Markovian process the time-scales are thus defined as the inverse of the eigenvalues of the FP operator. The methodology of eigenfunction expansion shows that the typology of time-scales included in a stochastic process is therefore strictly linked to the auto correlation properties of the process itself. The generalization to a multiplicative stationary Markovian process has been considered in Ref. \cite{paperpa19, paperpa20}, where we considered stationary Markovian processes that include multiple time-scales  \cite{GS1,SK,GS2,HL,Zoller,farago}. In fact, we considered explicit examples of $(i)$ short-range correlated processes with an autocorrelation function that shows a decay modulated by some (stretched-) exponential function, $(ii)$ short-range correlated processes with an autocorrelation function that shows a power-law decay and $(iii)$ long-range correlated processes, thus showing anomalous diffusion \cite{Beran94, anomalous1, anomalous2,Oliveira,Chavanis}. 

By using such explicit examples of stationary Markovian stochastic processes, the investigations performed in Ref. \cite{paperpa19, paperpa20} mainly regarded the relationship between the stationary pdf or the conditional probability on one side and on the other side the typology of time-scales included in the process, i.e. its correlation properties. The aim of this paper is to investigate how the inclusion of more and more time-scales into stationary Markovian stochastic processes affects the First Passage Time distribution (FPTD). 

First Passage Time issues are a classical topic in stochastic proceses research \cite{FPT1, FPT2}. They also have relevant applications, for example, in many fields of finance. Just to mention, the assessment of the default risk for firms' assets is essentially based on the evaluation of the FPT under the assumption that the asset value $V$ follows a geometric brownian motion \cite{finance1, finance2, finance3}, i.e. an Ornstein-Uhlembeck process for the variable $x=\log(V)$. There has therefore been much effort in obtaining the FPTD of the OU process \cite{OU1, OU2, OU4} thus neglecting the possible implications of the correlation properties of the asset value at the level of the FPTD and therefore at the level of the risk assessment. 

In this paper we will start considering the effects of the autocorrelation properties of a stationary Markovian stochastic process on the FPTD. Deliberately, our investigations will be at a theoretical level, with the aim of understanding what are the relevant aspects of the problem. To this end we will concentrate on the First Passage Time through two barriers located at $x=a$ and $x=b$ starting from position $x_0 \in [a,b]$. In fact, as explained below, such problem can be treated analytically. The search for a more realistic model in the context of default risk or for other financial variables \cite{MET} will be left for the future.

In this paper we will show that the large time dependance of the FPTD crucially depends on the correlation properties of the process and it is independent from its stationary distribution. Moreover, in the case when the process is power-law correlated the FPTD  shows power-law tails $1/t^{(\alpha+1)/2}$. As a result, in the case when the process is power-law correlated the moments $\langle t^n \rangle$ of the FPTD are finite only when $n< (\alpha-1)/2$. Finally,  we will give a general result connecting the FPTD of an additive stochastic processes $x(t)$ to the FPTD of a generic process $y(t)$ related by a coordinate transformation $y=f(x)$ to the first one.

The paper is organized as follows. In section \ref{proc} we will give explicit examples of stationary Markovian processes that include multiple time-scales. We will consider short-range correlated processes that include an infinite set of time-scales bounded from above and processes that include an infinite set of time-scales that extends up to infinity. For both cases we will consider the case when the stationary pdf is gaussian or with power-law tails. In section \ref{MFPT} we will consider the mean FPT for the processes considered in section \ref{proc}. We will show how the simple knowledge of the mean FPT may be not enough to let us discern between processes with different correlation properties. In section \ref{FPTD} we will derive a general expression for the FPTD of stationary Markovian stochastic processes. By using the explicit examples of section \ref{proc} we will then show how the correlation properties of the considered process affect the tails of the FPTD. Finally, in section \ref{concl} we will draw our conclusions.
 
\section{Processes }   \label{proc}

In this section we will briefly illustrate the stochastic processes to be considered in this paper. Further details can be found in Ref. \cite{paperpa19} and Ref. \cite{paperpa20}.

Let us consider stationary Markovian processes that can be described by a FP equation \cite{risken}:
\begin{eqnarray}
                               {\partial \over {\partial t}} W(x,t)  = - {\partial\over {\partial x}} \bigl( h(x) W(x,t) \bigl)  + {\partial^2 \over {\partial x^2}} \bigl( g(x)^2 W(x,t) \bigl) \label{FPE}
\end{eqnarray}
where $h(x)$ and $g(x)$ are the drift and diffusion coefficient appearing in the associated Langevin equation:
\begin{eqnarray}
                              \dot{x}(t)= h(x)\, + g(x)\,\Gamma(t)  \label{lang}
\end{eqnarray}
where $\Gamma(t)$ is a $\delta$--correlated Gaussian noise term. 

For a stochastic process with a constant diffusion coefficient, the eigenvalue spectrum of the FP equation describing a stationary process consists of a discrete part  ${\lambda_0=0,\lambda_1,...,\lambda_p}$ and a continuous part  $]\lambda_c,+\infty[$  ($\lambda_c \geqslant \lambda_p$)  associated with eigenfunctions $\varphi_{\lambda}$. The FP equation with constant diffusion coefficient can be transformed into a Schr\"odinger equation \cite{risken} with a quantum potential $V_S(x)=h(x)^2/4+\partial_x h(x)/2$. The eigenvalue spectrum of the Schr\"odinger equation is equal to the eigenvalue spectrum of the FP equation. The relation between the eigenfunctions of the FP equation and the eigenfunctions $\psi_{\lambda}$ of the Schr\"odinger equationis $\varphi_{\lambda}=\psi_{\lambda}\psi_0$. 

For a stationary process such that $V_S(x)$ only admits a null eigenvalue and a continuum part of the spectrum, one can show that:
\begin{equation} 
                \langle x(t+\tau)x(t)\rangle -  \langle x(t)\rangle  \langle x(t+\tau)\rangle=\int_{\lambda_c}^{+\infty}~c^2_{\lambda}e^{-\lambda \tau} d\lambda, \label{COV}
\end{equation}
where $c_{\lambda} \equiv \int_{-\infty}^{\infty} dx~x\,\varphi_{\lambda}(x)$. This shows that the autocovariance function is the infinite weighted sum of exponential functions $e^{-\lambda t}$ each characterized by a time-scale $\lambda^{-1}$, where the $\lambda$ values are the eigenvalues of the FP operator.
 
Eq. $(\ref{COV})$ clearly shows that the typology of time-scales included in a stochastic process is strictly linked to the auto correlation properties of the process itself. A generalization to a multiplicative stationary Markovian process has been considered in Ref. \cite{paperpa19, paperpa20}.

\subsection{A short-range correlated process}   \label{risk}

Let us consider the stochastic process described by the following Langevin equation \cite{risken}:
\begin{eqnarray}
               && \dot{x}(t)=- h(x(t))\, + D\,\Gamma(t),  \nonumber \\
               &&  h(x)=\left \{ \begin{array}{cc}
                                +k   &{\rm{if}}~~x < 0 ,\\
                                     &   \\
                                -k   &{\rm{if}}~~x >  0, 
                        \end{array} \right. \label{D1risk}
\end{eqnarray}
where $k$ is a real constant and $\Gamma(t)$ is a $\delta$--correlated gaussian noise term. The diffusion coefficient $D$ will be set to unity hereafter. It is straightforward to show that the stationary distribution of the process is:
\begin{eqnarray}
               W_s(x)={k\over 2}\,{\rm{exp}}( -{k\over 2} |x|). \label{pdfrisken}
\end{eqnarray}
The above process is a stochastic Markov process described by a FP equation with constant diffusion coefficient. As such, the FP equation maps onto an equivalent Schr\"odinger equation \cite{risken} with an appropriate quantum potential $V_S(x)$. In this case the potential is given by: $V_S(x)$$=k^2/4 - k\,\delta(x)$. It admits a ground state with null eigenvalue and a continuum set of eigenvalues $\lambda> k^2/4$ whose eigenfunctions are given in Ref. \cite{risken}.  By using the methodology of eigenfunction expansion \cite{risken,gardiner} it is possible to prove that the autocovariance function $R_s(t)=\langle x(t)x(0)\rangle$ of the above process is:
\begin{eqnarray}
          &&   R_s(t)={2 \over k^2} (1 - 2 \tau+4 \tau^2+ {8 \over 3} \tau^3)\,\bigl(1-{\rm{Erf}}(\sqrt{\tau}) \bigl)+  \label{ACrisken} \\
          &&   \quad -\,{{4 \sqrt{\tau}}\over{3 k^2 \sqrt{\pi}}}\,(2 \tau-1)(3 + 2 \tau)\,{\rm{exp}}(- \tau),                                 \qquad 
               \tau={k^2 \over 4} t, \nonumber
\end{eqnarray}
which behaves like a power-law with an exponential truncation for large time lags: $R_s(t) \approx {\rm{exp}}(-\tau) \tau^{-3/2}$ as $t \to \infty$. As a result, this is a short-range correlated process characterized by an infinite set of time-scales bounded from above, i.e. $\lambda^{-1}< 4/k^2$.

\subsection{A power-law correlated process}   \label{chi}

Let us consider the stochastic process described by the following nonlinear Langevin equation \cite{Zoller,paperpa19}:
\begin{eqnarray}
          &&  \dot{x}=h(x(t))+ \Gamma(t), \nonumber \\ 
          &&  h(x)=\left \{ \begin{array}{cc}
               -2 \sqrt{V_0} \tan (\sqrt{V_0}x) &{\rm{if}}~~|x| \leq L ,\\
                                         & \\
                 (1-\sqrt{1+4~V_1})/ x   &{\rm{if}}~~|x| >   L ,   
                                  \end{array} \right., \label{D1chimera} \\
          &&  V_1=L  \sqrt{V_0} \tan\bigl( \sqrt{V_0} L\bigl) \Bigl(1+L  \sqrt{V_0} \tan\bigl( \sqrt{V_0} L\bigl)\Bigl)   \nonumber                 
\end{eqnarray}
where $L$ and $V_0$ are real arbitrary constants and $\Gamma(t)$ is a $\delta$--correlated gaussian noise term. It is straightforward to show that the stationary distribution of this process is:
\begin{eqnarray}
            &&  W_l(x)=\left \{ \begin{array}{cc}
                                   N_{II} \cos(\sqrt{V_0}x) &{\rm{if}}~~|x| \leq L ,\\
                                                                             &   \\
                                   N_I 1/x^\alpha   &{\rm{if}}~~|x| >   L ,   
                              \end{array} \right. \label{pdfCHI} \\
            && \alpha=\sqrt{1+4\,V_1}-1 = 2 L  \sqrt{V_0} \tan\bigl( \sqrt{V_0} L\bigl),     \nonumber                
\end{eqnarray}
where $N_I$ and $N_{II}$ are real constants that can be obtained by imposing that $W_l(x)$ is continuous and normalized to unity. 

The  quantum potential associated with this process is given by: 
\begin{eqnarray}
              V_S(x)=\left \{\begin{array}{ccc}
                                   -V_0~    &~{\rm{if}}~&~|x| \leq L, \\
                                            &           &                  \\
                                   V_1/x^2 ~&~{\rm{if}}~&~|x| >   L,
                   \end{array} \right.  \label{VSchimera}
\end{eqnarray} 
Such potential admits a ground state with null eigenvalue and a continuum set of eigenvalues $\lambda> 0$ attached to the null eigenvalue. 

The eigenfunction of the ground state is
$\psi_0=B_0 \,\cos(\sqrt{V_0}\,x)$ whereas for $|x|>L$ it decays according to $\psi_0=A_0 \,x^{-\alpha/2}$.
The constants $A_0$ and $B_0$ are set by imposing that $\psi_0$ is normalized and continuous in $x=\pm L$. It is worth noting that for $|x|> L$ the stationary pdf $W(x)$ of the stochastic process is a power-law function decaying as $|x|^{-\alpha}$. The normalizability of the eigenfunction of the ground state is ensured if $\alpha>1$.
In the present study we consider stochastic processes with finite variance which implies $\alpha > 3$. For $|x| > L$ the eigenfunction $\psi_\lambda$ is a linear combination of Bessel functions $\psi_\lambda=A_\lambda\,\sqrt{x} J_\nu(\sqrt{\lambda}\,x)+B_\lambda\,\sqrt{x}\, Y_\nu(\sqrt{\lambda}\,x)$ where $\nu=(\alpha+1)/2$. For $|x| \leqslant L$ we find $\psi_\lambda^{(odd)}=D_\lambda\,\sin(\sqrt{V_0+\lambda}\,x)$ and $\psi_\lambda^{(even)}=F_\lambda\,\cos(\sqrt{V_0+\lambda}\,x)$. The coefficients $A_\lambda$, $B_\lambda$, $D_\lambda$ and $F_\lambda$ are fixed by imposing that $\psi_\lambda$ and its first derivative are continuous in $x=\pm L$ and that $\psi_\lambda$ are orthonormalized with a $\delta$-function of the energy.

By using the methodology of eigenfunction expansion \cite{risken,gardiner} it is possible to prove that the autocovariance function $\langle x(t+\tau)\,x(t) \rangle \propto \tau^{-\beta}$, where $\beta=(\alpha-3)/2$ thus showing that we are dealing with a power-law correlated stochastic process. In the range $3<\alpha<5$ the process is long-range correlated. As a result, this is a power-law correlated process admitting an infinite and unbounded set of time-scales.

\subsection{A short-range correlated process with a gaussian pdf}   \label{riskgaussall}

Let us now consider the process of section \ref{risk} and the coordinate transformation \cite{paperpa20}:
\begin{eqnarray}
                 x \mapsto y=f_g(x)=\sqrt{2\,s}\,{\rm{Erf}}^{-1}\bigl[1-e^{- k |x|}\bigr]  \label{risken_to_gauss}
\end{eqnarray}
By using the Ito lemma, one can show that, starting from the process of Eq. $(\ref{D1risk})$, in the coordinate space $y=f_g(x)$ one gets a multiplicative stochastic process whose stationary pdf is given by:
\begin{eqnarray}
               W_g(y)={1\over{\sqrt{2 \pi s}}} {\rm{exp}}(-{1 \over{2 s}} y^2)  \label{pdfOU}
\end{eqnarray}
with $s$ an additional arbitrary parameter. The corresponding drift and diffusion coefficients are given by:
\begin{eqnarray}
                       &&       H(y)=  {1 \over 2}\,k^2\,e^{y^2/(2 s)}\,\, \Bigl(1-{\rm{Erf}}\Bigl(|y|/\sqrt{2 s}\Bigl) \Bigl)\,\,\times \label{hRGA}  \\
                       &&       \hspace{1 cm}            \times\,\Bigl(
                                                                     \pi\,e^{y^2/(2 s)}\,\Bigl(1-{\rm{Erf}}\Bigl(|y|/\sqrt{2 s}\Bigl) \Bigl) - \epsilon\,2 \sqrt{2 \pi s}
                                                             \Bigl)   \nonumber \\
                       &&       G(y)=  \sqrt{\pi \over 2}\,k\,\sqrt{s}\,e^{y^2/(2 s)}\, \Bigl(1-{\rm{Erf}}\Bigl(|y|/\sqrt{2 s}\Bigl) \Bigl)\label{gRGA} 
\end{eqnarray}
where $\epsilon=+1$ when $y>0$ and $\epsilon=-1$ when $y<0$. For large values of $y$ the drift and diffusion coefficient behave as:
\begin{eqnarray}
              H(y) \propto - k^2 s\,{1 \over y} 
                           \qquad 
              G(y) \propto + k\,s\,{1 \over y} 
                           \qquad 
              y \to + \infty. \nonumber 
\end{eqnarray}

According to the eigenfunction expansion methodology \cite{risken,gardiner}, the autocorrelation function of the process defined by Eq. $(\ref{D1risk})$ with the coordinate transformation of Eq. $(\ref{risken_to_gauss})$ is given by $\rho_g(\tau)= (\langle y(t+\tau)y(t)\rangle-\langle y(t)\rangle^2)/(\langle y^2(t)\rangle-\langle y(t)\rangle^2)$ where:
\begin{eqnarray}
       &&   \langle y(t) y(t+\tau) \rangle=\int_{\lambda_c}^\infty d \lambda\,{\cal{C}}_\lambda^{2} e^{- \lambda \tau},  \label{Rrisk}  \\
       &&   \lambda_c=k^2/4, \qquad {\cal{C}}_\lambda=\int_{-\infty}^{+\infty} dx\,f_g(x)\,\psi_0(x)\,\psi_\lambda(x) ,\nonumber
\end{eqnarray}
where $\psi_0(x)$ and $\psi_\lambda(x)$ are the eigenfunctions of the Schr\"odinger equation with potential $V_S(x)$$=k^2/4 - k\,\delta(x)$ associated to the stochastic process of Eq. $(\ref{D1risk})$. 

As explained in Ref. \cite{paperpa20} such process admits an infinite set of time-scales bounded from above and its autocorrelation function decays like $R_g(t) \approx {\rm{exp}}(- k^2/4 t) t^{-3/2}$ for large time lags. As a result, this is a gaussian distributed, short-range correlated process, characterized by an infinite set of time-scales bounded from above.

\subsection{A short-range correlated process with a power-law decaying pdf}   \label{RPA}

Let us now consider again the process described by Eq. $(\ref{D1risk})$. By performing the following coordinate transformation:
\begin{eqnarray}
                 x \mapsto y=f_p(x)=\left \{ \begin{array}{l}
                                                              (k\,L^\alpha)^{1/(\alpha-1)}\,e^{k x/(\alpha-1)},     \hfill           ~|x| > L/D, \\
                                                                                                   \\   
                                                              D x,     \hfill           ~|x| \le L/D,   
                                                     \end{array} \right.    \label{risken_to_pow}
\end{eqnarray}
and by using the Ito lemma, one can show that in the coordinate space $y=f_p(x)$ one gets a multiplicative stochastic process whose stationary pdf is given by:
\begin{eqnarray}
               W_p(y)=\left \{ \begin{array}{l}
                                                              {{k L^\alpha} \over {2 D}}\,e^{- k L/D}\,{1 \over y^\alpha},    \hfill           ~|y| > L, \\
                                                                                                   \\   
                                                              {{k} \over {2 D}}\, e^{- k y/D} ,   \hfill           ~|y| \le L  , 
                                                     \end{array} \right.  \label{pdfRPA}
\end{eqnarray}
with $L$ and $\alpha$ positive real constants. The requirement that the diffusion coefficient is continuous in $y = \pm L$ gives the following constraint:
\begin{eqnarray}
                               k=(\alpha-1)\,{{D}\over{L}}. \label{alfak}
\end{eqnarray}
Correspondingly, the drift and diffusion coefficients are given by:
\begin{eqnarray}
                       &&       H(y)= \left \{ \begin{array}{l}
                                                              -(\alpha-2)/L^2\,D^2\,y,    \hfill           ~|y| > L, \\
                                                                                                   \\   
                                                               +(\alpha-1)/L\,D^2, \hfill           ~|y| \le L,   
                                                     \end{array} \right.  \label{hRPA}  \\
                       &&       G(y)=  \left \{ \begin{array}{l}
                                                              D/L |y|,    \hfill           ~|y| > L, \\
                                                                                                   \\   
                                                              D,  \hfill           ~|y| \le L,   
                                                     \end{array} \right.    \label{gRPA} 
\end{eqnarray}

According to the eigenfunction expansion methodology \cite{risken,gardiner}, the autocorrelation function of the process defined by Eq. $(\ref{D1risk})$ with the coordinate transformation of Eq. $(\ref{risken_to_pow})$ is given by Eq. $(\ref{Rrisk})$ with $f_g(x)$  replaced by $f_p(x)$. The relevant integrations can be performed analytically. One can therefore show that such process admits an infinite set of time-scales bounded from above and its autocorrelation function decays like $R_p(t) \approx {\rm{exp}}(-k^2/4 t) t^{-3/2}$ for large time lags. As a result, this is a short-range correlated process with a power-law decaying stationary pdf, characterized by an infinite set of time-scales bounded from above.

\subsection{A power-law correlated process with a gaussian pdf}   \label{chigaussall}

Let us now consider the process of section \ref{chi} and the coordinate transformation \cite{paperpa20}:
\begin{eqnarray}
          &&  x \mapsto y=f_l(x)= \left \{ \begin{array}{l}
                                f_I=\sqrt{2\,s}~{\rm{Erf}}^{^{-1}}
                                       \bigl[1-r(x) \bigl] ,    \hfill           ~|x| > L, \\
                                                                                                   \\   
                                f_{II}=\sqrt{2\,s}~{\rm{Erf}}^{^{-1}}
                                       \bigl[1-p(x) \bigl] ,    \hfill           ~|x| \le L  , 
                       \end{array} \right.   \nonumber  \\
          &&    r(x)={
                      {2\,N_I\,x^{1-\alpha}}
                      \over
                      {\alpha-1}
                     }   ,  \label{CHI_to_gauss}     \\
          &&    p(x)={{2\,N_I\,L} \over {L^\alpha} }-
                     {{N_I}\over{L^\alpha\,\cos(\sqrt{V_0} L)^2}}\,\bigl(x - L\bigr)+ \nonumber \\
          &&       -~{{N_I}\over{2 \sqrt{V_0}\,L^\alpha\,\cos(\sqrt{V_0} L)^2}} 
                     \Bigl(\sin(2 \sqrt{V_0} x)-  \sin(2 \sqrt{V_0} L)\Bigr) . \nonumber
\end{eqnarray}
By using the Ito lemma, one can show that, starting from the process of Eq. $(\ref{D1chimera})$, in the coordinate space $y=f_l(x)$ one gets a multiplicative stochastic process whose stationary pdf is given by Eq. $(\ref{pdfOU})$, with $s$ as an additional arbitrary parameter. One can easily show that $G(y)$ is continuos along the real axis, although its first derivative is discontinuos in $y=\pm f(L)$. The drift coefficient $H(y)$ is discontinuos in $y=\pm f(L)$. For large values of $y$ the drift and diffusion coefficient behave as:
\begin{eqnarray}
             && H(y) \propto {1 \over{y^{{\alpha+1}\over{\alpha-1}}}} \exp\bigl({-{y^2\over{s (\alpha-1)}}}\bigr), \qquad y \to +\infty, \nonumber \\
             && G(y) \propto {1 \over{y^{{\alpha}\over{\alpha-1}}}} \exp\bigl({-{y^2\over{2 s (\alpha-1)}}}\bigr), \qquad y \to +\infty. \nonumber
\end{eqnarray}

According to the eigenfunction expansion methodology \cite{risken,gardiner}, the autocorrelation function of the process defined by Eq. $(\ref{D1chimera})$ with the coordinate transformation of Eq. $(\ref{CHI_to_gauss})$ is given by Eq. $(\ref{Rrisk})$ with $f_g(x)$ replaced by $f_l(x)$, $\lambda_c=0$ and the eigenfunctions $\psi_0$ and $\psi_\lambda$ are solutions of the Schr\"odinger equation with the quantum potential of Eq. $(\ref{VSchimera})$. 

As explained in Ref. \cite{paperpa19} and Ref. \cite{paperpa20} such process admits an infinite and unbounded set of time-scales and its autocorrelation function decays like $R_l(t) \approx 1/t^{(\alpha-1)/2}$ with logarithmic corrections for large time lags. As a result, this is a gaussian distributed, power-law correlated process, characterized by an infinite and unbounded set of time-scales.

\section{The Mean First Passage Time}   \label{MFPT}

In order to understand how the correlation properties of a stochastic process affect its FPTD we start investigating the mean FPT, which is the first moment of the FPTD. We will show in this section that processes with different correlation properties might show a similar behaviour in the  mean FPT. 

Let us  consider the mean time $T_{x}(\Lambda)$ that is needed to reach for the first time position $x \pm \Lambda$ starting from position $x$.  This is obtained by solving the equation \cite{gardiner}:
\begin{eqnarray}
                             h(x) { \partial T_{x}(\Lambda) \over \partial x} + g(x)^2 { \partial^2 T_{x}(\Lambda) \over \partial x^2} = -1 \label{FPTequation}
\end{eqnarray}
with boundary conditions $T_{x}(\Lambda)=0$ when $x=\pm \Lambda$. Here $h(x)$ and $g(x)$ are the drift and diffusion coefficient appearing in the non-linear Langevin equation associated to the considered stochastic process.

\subsection{Explicit examples of processes with a gaussian stationary distribution}   \label{gausspropagator}

In this section we will compare the mean FPT of the two stochastic processes with gaussian pdf considered in the previous section. Indeed, the results shown here have already been considered in Ref. \cite{paperpa20}. However, for the sake of clarity we present them here again in a shortened form.

For the stochastic gaussian process of section \ref{riskgaussall} the mean FPT can be analytically computed by using the drift coefficient of Eq. $(\ref{hRGA})$ and the diffusion coefficient of Eq. $(\ref{gRGA})$ into Eq. $(\ref{FPTequation})$. For the case when $x=0$ one gets:
\begin{eqnarray}
                      &&       T_0(\Lambda)= {1 \over k^2} \,\log\Bigl(1- {\rm{Erf}}\bigl(\Lambda/\sqrt{2 s}\bigl) \Bigl)\,+\label{FPTRGA} \\
                      &&        \hspace{2.5 truecm} -\,{1 \over k^2}\Bigl(1- {1 \over{1- {\rm{Erf}}\bigl(\Lambda/\sqrt{2 s}\bigl)}}\Bigl). \nonumber
\end{eqnarray}
An alternative way to obtain such result is to analytically compute the mean FPT for the stochastic process of Eq. $(\ref{D1risk})$ and then performing the substitution:
\begin{eqnarray}
                      \Lambda \mapsto - {1 \over k} \log\Bigl(1- {\rm{Erf}}\bigl(\Lambda/\sqrt{2 s}\bigl) \Bigl).
\end{eqnarray}
For large values of $\Lambda$ the mean FPT of Eq. $(\ref{FPTRGA})$ gives:
\begin{eqnarray}
                      &&       T_0(\Lambda) \approx z \, e^{z^2}, \qquad z={\Lambda \over \sqrt{2 s}}. \label{FPTRGAas}
\end{eqnarray}

For the process of section \ref{chigaussall} the mean FPT can not be computed by analytically solving Eq. $(\ref{FPTequation})$. However, one can follow the alternative way mentioned above. Specifically, we first computed analytically the mean first passage time of the process of Eq. $(\ref{D1chimera})$ by using Eq. $(\ref{FPTequation})$. Subsequently, according to Eq. $(\ref{CHI_to_gauss})$, we performed the substitution:
\begin{eqnarray}
                      &&    \Lambda \mapsto f_I^{-1}(\Lambda),        \qquad \qquad {\rm{if}}  \,  \Lambda>L, \\
                      &&    \Lambda \mapsto f_{II}^{-1}(\Lambda),     \qquad \qquad {\rm{if}}  \,  \Lambda<L.
\end{eqnarray}
In the region $\Lambda>f_I(L)$ it is possible to obtain an analytical expression showing that for large values of $\Lambda$ one gets:
\begin{eqnarray}
                      &&       T_0(\Lambda) \approx \Bigl(  z \, e^{z^2} \Bigl)^{{\alpha+1}\over{\alpha-1}}, \qquad z={\Lambda \over \sqrt{2 s}}. \label{FPTCGAas} 
\end{eqnarray}
Eq. $(\ref{FPTCGAas})$ and Eq. $(\ref{FPTRGAas})$ clearly show that for large values of $\Lambda$ the process of section \ref{chigaussall} has a mean FPT larger than the process of section \ref{riskgaussall}. This confirms that the process that incorporates more time-scales is the slowest.

In the two examples considered here, the way the correlation properties of the process affect the mean FPT is clear. In the case of Eq. $(\ref{FPTRGAas})$ we have a linear growth in terms of the $e^{z^2}$ function, while in the case of Eq. $(\ref{FPTCGAas})$ we have a power-law growth with exponent larger than unity.

\subsection{Explicit examples of processes with a power-law decaying stationary distribution}   \label{powerpropagator}

Let us now consider the mean FPT of the two stochastic processes with power-law decaying pdf considered in section \ref{proc}. 

For the stochastic process of setion \ref{RPA} the mean first passage time can be analytically computed by using the drift coefficient and the diffusion coefficient of Eq. $(\ref{hRPA})$ and Eq. $(\ref{gRPA})$ into Eq. $(\ref{FPTequation})$. It is possible to obtain an analytical expression showing that for large values of $\Lambda$ one gets:
\begin{eqnarray}
                      &&       T_0(\Lambda) \propto \Lambda^{\alpha-1}, \label{FPTrpa} 
\end{eqnarray}
where $\alpha$ is defined in Eq. $(\ref{alfak})$. For small values of $\Lambda$ one gets $T_0(\Lambda) \approx \Lambda^2/2$.

As mentioned in the previous section, for the stochastic process of section \ref{chi} the mean FPT can be analytically computed by using the drift coefficient and the diffusion coefficient of Eq. $(\ref{D1chimera})$ into Eq. $(\ref{FPTequation})$. It is possible to obtain an analytical expression showing that for large values of $\Lambda$ one gets:
\begin{eqnarray}
                      &&       T_0(\Lambda) \propto \Lambda^{\alpha+1}. \label{FPTOUmultas} 
\end{eqnarray}
For small values of $\Lambda$ one gets $T_0(\Lambda) \approx \Lambda^2/2$.

Both in Eq. $(\ref{FPTrpa})$ and $(\ref{FPTOUmultas})$ the parameter $\alpha$ is the exponent of the pdf tail. Therefore, as for the processes with gaussian pdf, the mean FPT of the process that includes more time-scales is larger than the other. However, in the two cases considered here a power-law dependance from $\Lambda$ is found. The qualitative behaviour of the mean FPT in both processes is therefore very similar, although the correlation properties of the two processes are very different. In fact, we will show that the different correlation properties of the stochastic processes display their effect at the level of the higher order FPTD moments.


\section{The First Passage Time Distribution}   \label{FPTD}

\subsection{A Theoretical result}   \label{theory}

Let us suppose to be interested in the distribution ${\cal{D}}_{a,b}(x_0,t)$ of the first times at which a process reaches the absorbing barriers $x=a$ or $x=b$ starting from a generic position $x_0$ with  $a<x_0<b$. According to Ref. \cite{gardiner} one gets:
\begin{eqnarray}
                      &&          {\cal{D}}_{a,b}(x_0,t)= \partial_t G_{a,b}(x_0,t), \nonumber \\
                      &&          G_{a,b}(x_0,t)=\int_a^b dx P(x,t|x_0,0),
\end{eqnarray}
where $P(x,t|x_0,0)$ is the conditional probability of finding the process in position $x$ at time $t$, knowing that it was in position $x_0$ at time $t=0$. According to the methodology of eigenfunction expansion \cite{gardiner,risken}, for an additive stationary Markovian process described by a FPE admitting only a continuum set of eigenvalues, the  conditional probability is given by:
\begin{eqnarray}
                                P(x,t|x_0,0)= \psi_0(x)^2 + { \psi_0(x) \over \psi_0(x_0)} \int_{\lambda_c}^{\infty} d \lambda e^{- \lambda t} \psi_\lambda(x)\,\psi_\lambda(x_0),
\end{eqnarray}
where the $\lambda$-integration is extended over the continuum part of the spectrum and $\psi_0(x)$, $\psi_\lambda(x)$ are the eigenfunctions of the quantum potential $V_S$ associated to the stochastic process considered. If the potential $V_S(x)$ admits a discrete part of the spectrum then an additional term involving a sum over all the discrete eigenvalues should be considered. By using the above expression for the conditional probability one finally gets:
 \begin{eqnarray}
                      &&         G_{a,b}(x_0,t) =  { 1 \over \psi_0(x_0)} \,\int_{\lambda_c}^\infty d \lambda \, e^{- \lambda t} \,\psi_\lambda(x_0) \times \nonumber \\
                      &&          \hspace{2.0 truecm}  \times \int_a^b d x  \, \psi_0(x) \psi_\lambda(x).  \label{FPTDx}
\end{eqnarray}
The role played by each time-scale in the FPTD can be emphasized by rewriting Eq. $(\ref{FPTDx})$ in the form
\begin{eqnarray}
                      &&         G_{a,b}(x_0,t) =  \int_{\lambda_c}^\infty d \lambda \, e^{- \lambda t} \, {\cal{W}}_\lambda(x_0;a,b), \nonumber \\
                      &&         {\cal{W}}_\lambda(x_0;a,b)={\psi_\lambda(x_0) \over  \psi_0(x_0)}\,\int_{a}^{b} d x  \, \psi_0(x) \psi_\lambda(x), \label{FPTDadd}
 \end{eqnarray}
where ${\cal{W}}_\lambda(x_0;a,b)$ can be considered as the weight with which each time-scale enters the FPTD.

The above formulae can be used to obtain the FPTD in the case of the additive stochastic processes of section \ref{risk} and section \ref{chi}. 

Let us now consider a stochastic process $y(t)$ obtained by performing a coordinate transformation $y=f(x)$ starting from a stochastic additive process $x(t)$, as for the processes of section \ref{riskgaussall}, section \ref{RPA} and section \ref{chigaussall}. The conditional probability is given by \cite{paperpa20}:
\begin{eqnarray}
                   &&       P(y,t|y_0,0)=W(y)+ \label{ProbCondEigen}\\
                   &&       +\, {1 \over {\partial f(x)}}\,
                                 {{\psi_0(x)} \over {\psi_0(x_0)}}\,
                                 \int_{\lambda_c}^{\infty} d\lambda \psi_\lambda(x)\,\psi_\lambda(x_0)\,e^{-\lambda t}, \nonumber \\
                   &&       x=f^{-1}(y), \qquad \qquad x_0=f^{-1}(y_0), \nonumber
\end{eqnarray}
where $W(y)$ is the stationary distribution of the process $y(t)$ and $\psi_0(x)$, $\psi_\lambda(x)$ are the eigenfunctions of the Schr\"odinger equation with the potential $V_S(x)$ associated to the additive process $x(t)$. As a result, the FPTD of the transformed process $y(t)$  is given by:
\begin{eqnarray}
                      &&         {\cal{D}}_{a,b}^f(y_0,t) =  { 1 \over \psi_0(x_0)} {\partial \over \partial t}\, \int_{\lambda_c}^\infty d \lambda e^{- \lambda t}  \,\psi_\lambda(x_0) \times \nonumber \\
                      &&          \hspace{2.0 truecm}  \times \int_{f^{-1}(a)}^{f^{-1}(b)} dx \, \psi_0(x)\psi_\lambda(x), \label{FPTDy} \\
                      &&          x_0=f^{-1}(y_0). \nonumber
\end{eqnarray}
A direct comparison of Eq. $(\ref{FPTDy})$ and Eq. $(\ref{FPTDx})$ shows that:
\begin{eqnarray}
                               {\cal{D}}_{a,b}^f(y_0,t) = {\cal{D}}_{f^{-1}(a),f^{-1}(b)}(f^{-1}(y_0),t). \label{additive_to_transf}
\end{eqnarray}
Therefore, for a general stochastic process $y(t)$, the problem of finding the FPTD can always be re-conducted to the problem of finding the FPTD of an appropriate additive process. When the additive process admits a quantum potential $V_S(x)$ that is exactly solvable then Eq. $(\ref{FPTDadd})$ can be used. If $V_S(x)$ is not exactly solvable, one might either take advantage of the approximations schemes widely used in quantum mechanics to estimate the eigenfunctions or revert to numerical simulations of the process to estimate the FPTD.

\subsection{Explicit examples}   \label{FPTDex}


Let us now consider the stochastic process of section \ref{risk}. Moreover, let us fix $x_0=0$ and $b=\Lambda$, $a=-\Lambda$ as in section \ref{MFPT}. The integration over $x$ required in Eq. $(\ref{FPTDx})$ can be performed analytically. It gives that ${\cal{W}}_\lambda(0;-\Lambda,\Lambda) \approx \sqrt{\lambda - k^2/4}$ when $\lambda \to k^2/4$. Therefore the  large time behaviour of the FPTD is given by \cite{Olver74}:
 \begin{eqnarray}
                      &&         {\cal{D}}_{-\Lambda,\Lambda}(0,t) \sim e^{-{k^2 \over 4}\,t}/t^{3/2}, \qquad t \to \infty.  \label{FPTDtailrisk}
\end{eqnarray}
According to the result of Eq. $(\ref{additive_to_transf})$, the same asymptotic behaviour is shared by the FPTD of the process of section \ref{riskgaussall} and section \ref{RPA}, although they have different pdfs.

\begin{figure} 
\begin{center}
              \resizebox{1\columnwidth}{!}{\includegraphics[scale=0.30]{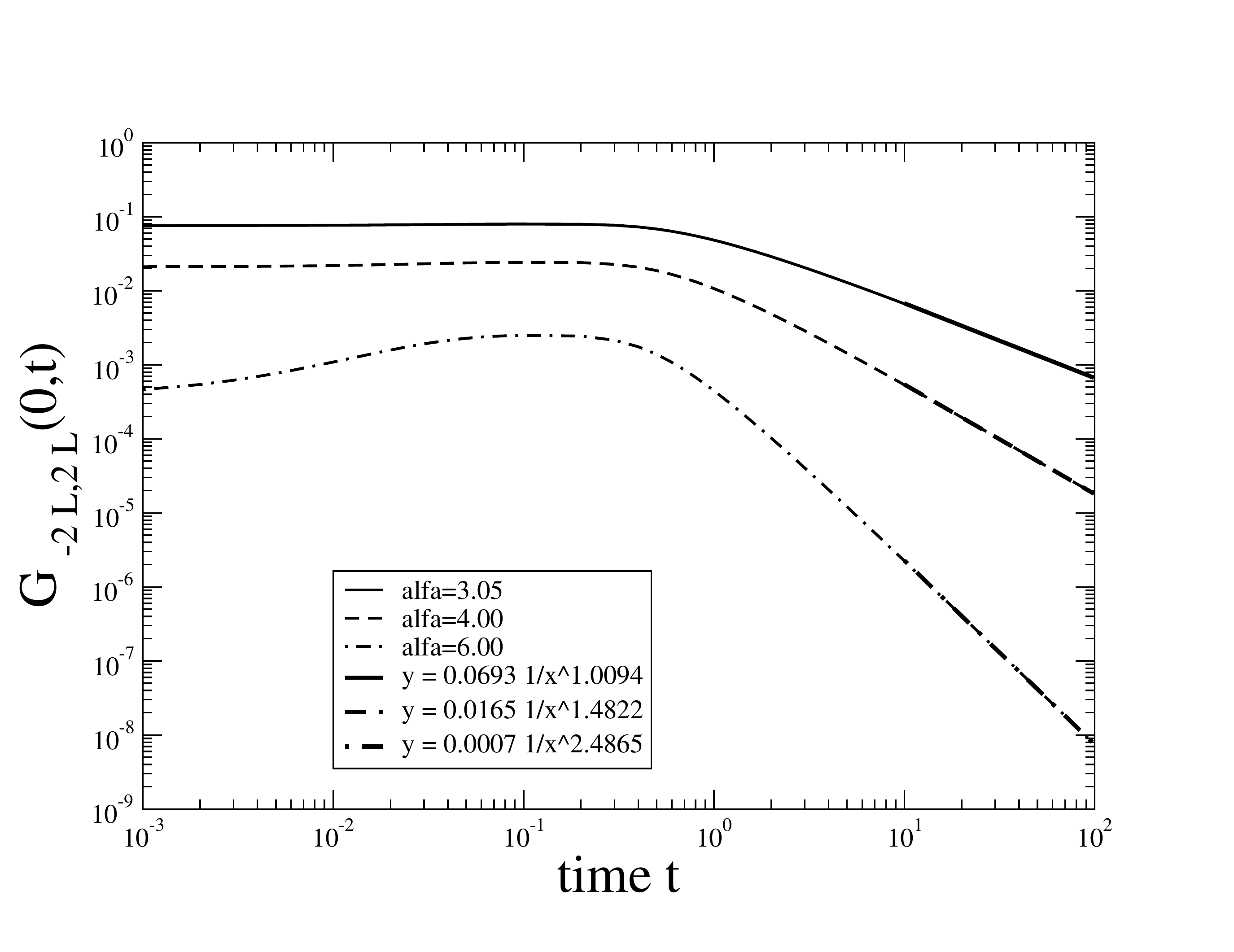} }
              \caption{The figure shows the function $G_{-\Lambda,\Lambda}(0,t)$ for the process of section \ref{chi}. The curves are obtained by using Eq. $(\ref{FPTDadd})$. We have considered the following three set of parameters: (i) $L=1$ and $V_0=0.985$ (solid line) corresponding to $\alpha=3.05$; (ii) $L=1$ and $V_0=1.16$ (dashed line) corresponding to $\alpha=4.00$; (iii) $L=1$ and $V_0=1.42272$ (dashed dotted line) corresponding to $\alpha=6.00$. For all the three sets we considered $\Lambda=2 L$.  In the bottom part of the legend we report the result of a nonlinear fit performed in the range $t \in[20,100]$. These results are clearly in agreement with those of Eq. $(\ref{FPTDtailchi})$.}  \label{FPTDintCHI}
\end{center}   
\end{figure}

Let us now consider the stochastic process of section \ref{chi} for the case when $x_0=0$ and $b=\Lambda$, $a=-\Lambda$. The integration over $x$ required in Eq. $(\ref{FPTDx})$ can be performed analytically. One therefore obtains that ${\cal{W}}_\lambda(0;-\Lambda,\Lambda) \approx \lambda^{(\alpha-3)/2}$. Therefore, the asymptotic behaviour of the FPTD is given by:
 \begin{eqnarray}
                      &&         {\cal{D}}_{-\Lambda,\Lambda}(0,t) \sim 1/t^{(\alpha+1)/2}, \qquad t \to \infty. \label{FPTDtailchi}
\end{eqnarray}
According to the result of Eq. $(\ref{additive_to_transf})$, the same asymptotic behaviour is shared by the FPTD of the process of section \ref{chigaussall}, although they have different pdfs.  In Fig. \ref{FPTDintCHI} we show the function $G_{-\Lambda,\Lambda}(0,t)$ for the process of section \ref{chi}. The curves are obtained by using Eq. $(\ref{FPTDadd})$. We have considered the following three set of parameters: (i) $L=1$ and $V_0=0.985$ (solid line) corresponding to $\alpha=3.05$; (ii) $L=1$ and $V_0=1.16$ (dashed line) corresponding to $\alpha=4.00$; (iii) $L=1$ and $V_0=1.42272$ (dashed dotted line) corresponding to $\alpha=6.00$. For all the three sets we considered $\Lambda=2 L$.  In the bottom part of the legend we report the result of a nonlinear fit performed in the range $t \in[20,100]$. These results are clearly in agreement with those of Eq. $(\ref{FPTDtailchi})$.

The above results explicitly show that the tail of the FPTD is independent from the specific stationary distribution of the stochastic process considered. Rather, it is strictly related to the correlation properties of the process. In fact the processes of section  \ref{chi} and section \ref{chigaussall} have the same FPTD tail although their pdf is different. Similarly for the three processes of section \ref{risk}, section \ref{riskgaussall} and section \ref{RPA}.

Another relevant comment regards the moments of the FPTD distribution. The existence of power-law tails in the FPTD of power-law correlated stationary Markovian processes implies that the moments of the distribution $\langle t^n \rangle$ are defined only for $n<(\alpha-1)/2$. In the case of the processes of section \ref{risk}, section \ref{riskgaussall} and section \ref{RPA}, the exponential cut-off of Eq. $(\ref{FPTDtailrisk})$ ensures that the moments of the FPTD are always finite. That marks a relevant difference between a process admitting a bounded set of timescales and a process in which all timescales are included. As shown in section \ref{powerpropagator}, such difference is not caught by looking at the mean FPT only. In fact both the processes of section \ref{RPA} and section \ref{chi}, both characterized by a power-law decaying pdf, have a power-law growing mean FPT. However, in the case of section \ref{RPA} all the higher moments of the FPTD exist, while in the case of section \ref{chi} only the moments $\langle t^n \rangle$ such that $n<(\alpha-1)/2$ are defined.

Finally, the result of Eq. $(\ref{FPTDtailchi})$ shows that the tail of the FPTD for power-law correlated stationary Markovian processes is different from what is expected for other non-Markovian processes with anomalous diffusion. For instance, in Ref. \cite{Ding} it is shown that for a Fractional Brownian motion the tails of the FPTD decay like $t^{\gamma/2-2}$, where $\gamma \in [0,2]$ is defined as $\langle X(t)^2 \rangle \propto t^\gamma$ and $\langle X(t)^2 \rangle$ is the mean square displacement of the process.

\subsection{Another process with an infinite and unbounded set of timescales}   \label{FPTDchi}

The fact that the process  of section \ref{chi}  includes an infinite and unbounded set of time-scales is a necessary condition for the existence of power-law tails in its FPTD. This is ultimately due to the fact that the eigenvalues spectrum of the Schr\"odinger potential of Eq. $(\ref{VSchimera})$ is given by a continuum set of eigenvalues $\lambda>0$ attached to the null eigenvalue associated to the ground state and therefore it includes an infinite and unbounded set of time-scales.. One might therefore wander whether or not such power-law decay is merely due to the fact that the process incorporates an unbounded set of timescales or something else must be considered. 

In general, any Schr\"odinger potential that asymptotically decays like $1/x^\mu$ would give a stochastic process with an infinite and unbounded set of time-scales. In the case $\mu=1$, we will show here that the corresponding FPTD tail does not necessarily display a power-law decay.

Let us now consider the additive stochastic process associated to the Schr\"odinger potential \cite{paperpa20}:
\begin{eqnarray}
              V_S(x)=\left \{\begin{array}{ccc}
                                   -V_0~    &~{\rm{if}}~&~|x| \leq L, \\
                                            &           &                  \\
                                   V_1/|x| ~&~{\rm{if}}~&~|x| >   L,
                   \end{array} \right.  \label{VScoulombwell}
\end{eqnarray} 
where $L$, $V_0$ and $V_1$ are real positive constants. This quantum potential is exactly solvable. In the region $x>L$ the eigenfunctions are given by:
\begin{eqnarray}
             && \psi_0(x)= A_0~\sqrt{x}~K_1(2\,\sqrt{V_1}~\sqrt{x}), \\
             && \psi_\lambda(x)= x~e^{- i \sqrt{\lambda} x}~\Bigl(                           
                                                                                    A_\lambda\,F_{1,1}(1 - i {V_1 \over 2}~{1 \over \sqrt{\lambda}}, 2, 2 i  \sqrt{\lambda} x)+ \nonumber \\
             &&         \hspace{2.5 truecm}                    C_\lambda\,U(1 - i {V_1 \over 2}~{1 \over  \sqrt{\lambda}}, 2, 2 i  \sqrt{\lambda} x)
                                                                           \Bigl)+c.c.  ,       \nonumber 
\end{eqnarray}
where $K_1(\cdot)$ is the K-Bessel function of order $1$, $U(\cdot)$ is the confluent hypergeometric function and $F_{1,1}(\cdot)$ is the Kummer hypergeometric function \cite{abramowitz}. 

The normalization constant $A_0$ is fixed by imposing that the ground state is normalized to unity. The other normalization constants $A_\lambda$ and $C_\lambda$ are fixed by imposing that the odd and even eigenfunctions fulfill the normalization condition $\int dx\, \psi_\lambda(x) \psi_{\lambda'}(x)=\delta(\lambda-\lambda')$. The drift coefficient $h(x)$ aymptotically decays like $h(x) \approx 1/\sqrt{x}$.

As mentioned above, the large time behaviour of ${\cal{D}}_{-\Lambda,+\Lambda}(0,t) $  is determined by the small energy behaviour of ${\cal{W}}_\lambda(0;-\Lambda,\Lambda)$ \cite{Olver74}. One can explitely show that in the limit when $\lambda$ is small then:
\begin{eqnarray}
               && A_\lambda\,F_{1,1}(1 - i {V_1 \over 2}~{1 \over \sqrt{\lambda}}, 2, 2 i  \sqrt{\lambda} x) \approx e^{- {{3 \pi V_1} \over {4\,\sqrt{\lambda}}}} \, \sqrt{x} \, I_1(2 \sqrt{V_1} x), \nonumber \\
               && C_\lambda\,U_{1,1}(1 - i {V_1 \over 2}~{1 \over \sqrt{\lambda}}, 2, 2 i  \sqrt{\lambda} x) \approx e^{- {{\pi V_1} \over {\sqrt{\lambda}}}} \, \sqrt{x} \, K_1(2 \sqrt{V_1} x), \nonumber
\end{eqnarray}
where $I_1(\cdot)$and $K_1(\cdot)$ are Bessel functions of order $1$. As a result, the small energy limit of $\psi_\lambda$ is essentially determined by the small energy limit of $A_\lambda\,F_{1,1}(1 - i {V_1 \over 2}~{1 \over \sqrt{\lambda}}, 2, 2 i  \sqrt{\lambda} x)$. By making use of Eq. $(13.6.8)$ and Eq. $(14.1.7)$ of Ref. \cite{abramowitz} one gets:
\begin{eqnarray}
                 &&   {\cal{W}}_\lambda(0;-\Lambda, \Lambda)=A_0 \, \psi_\lambda(0) \,{A_\lambda \over \sqrt{\lambda} } 
                                                                                                     {{e^{{\pi V_1}\over{4 \sqrt{\lambda}}}} \over{|\Gamma(1+i {{V_1} \over {2 \sqrt{\lambda}}})|}}   \times  \\
                 && \hspace{1.0 truecm}  \times   \int_L^\Lambda dy \sqrt{y} K_1(2 \sqrt{V_1} y) \, F_0({{V_1} \over {2 \sqrt{\lambda}}}, \sqrt{\lambda} y) , \nonumber
 \end{eqnarray}
where $F_0$ is the regular Coulomb function with null index. By making use of Eq. $(14.4.1)$ of Ref. \cite{abramowitz} one gets:
\begin{eqnarray}
                 &&  {\cal{W}}_\lambda(0;-\Lambda, \Lambda)=A_0 \, V_1^{-1}\,A_\lambda \, \psi_\lambda(0) \times \nonumber \\
                 && \hspace{1.0 truecm}  \times   \int_L^ \Lambda dy \sqrt{y} K_1(2 \sqrt{V_1} y) \, \sum_{k=1}^{\infty} b_k I_k(2 \sqrt{V_1} y),  \nonumber
 \end{eqnarray}
where $I_k(\cdot)$ is the Bessel functions of order $k$ and the coefficients $b_k$ are defined in Eq. $(14.4.6)$ of Ref. \cite{abramowitz}. We have thus factorized the $\lambda$ dependencies from the $y$ dependencies. As a result, the large time behaviour of the FPTD is given by the small energy behaviour of $A_\lambda \, \psi_\lambda(0) \approx e^{-{3 \pi} \over{2 \sqrt{\lambda}}}$. By using the results of Ref. \cite{Olver74} one finally gets:
\begin{eqnarray}
                 &&   {\cal{D}}_{-\Lambda,\Lambda}(0,t)   \approx  {{e^{- \kappa t^{1/3}}} \over {t^{5/6}}} \qquad \kappa=3\,\Bigl( { { 3\pi \over 4} V_1} \Bigl)^{2/3}
 \end{eqnarray}
The above result shows that the mere existence of an infinite and unbounded set of timescales is not enough to ensure an asymptotic power-law behaviour in ${\cal{D}}_{-\Lambda,\Lambda}(0,t)$. In passing, by using the same arguments it is also possible to prove that the stochastic process associated to the quantum potential of Eq. $(\ref{VScoulombwell})$ is not power-law correlated, even though it incorporates an infinite and unbounded set of timescales \cite{farago}.

\section{Conclusions}   \label{concl}


The aim of this paper was to investigate how the FPTD is affected by the typology of time-scales included in a stochastic process, i.e. by the correlation properties of the process itself. The motivations are both theoretical and in view of possible applications. In fact, First Passage Time issues have relevant applications in many fields of finance such as the assessment of the default risk for firms' assets.

In this paper we have shown that the tail of the FPTD crucially depends on the correlation properties of the process and it is independent from its stationary distribution. When the process includes an infinite set of time-scales bounded from above the FPTD shows tails modulated by some exponential decay. In the case when the process is power-law correlated $\langle x(t) x(t+\tau) \rangle \sim \tau^{-(\alpha-3)/2}$ the FPTD  shows power-law tails $1/t^{(\alpha+1)/2}$. As a result, in the case when the process is power-law correlated the moments $\langle t^n \rangle$ of the FPTD are finite only when $n< (\alpha-1)/2$. We have also shown that such power-law behaviour is not merely due to the fact that the process includes an infinite and unbounded set of time-scales. Rather, it is also required that each time-scale enters the FPTD in a peculiar way. In fact, in order to observe a power-law tail in the FPTD the weights ${\cal{W}}_\lambda(x_0;a,b)$ must be distributed according to a power-law for small values of $\lambda$, i.e. for large time-scales.

Moreover, we have given a general result showing that given (i) and additive stochastic processes $x(t)$ and (ii) another process $y(t)$ related by a coordinate transformation $y=f(x)$ to the first one, the FPTD of $y(t)$ can be obtained from the one of the $x(t)$ process by using Eq. $(\ref{additive_to_transf})$. Therefore, for a general stochastic process $y(t)$, the problem of finding the FPTD can always be re-conducted to the problem of finding the FPTD of an appropriate additive process. When the additive process admits a quantum potential $V_S(x)$ that is exactly solvable then Eq. $(\ref{FPTDadd})$ can be used. If $V_S(x)$ is not exactly solvable, one might either take advantage of the approximations schemes widely used in quantum mechanics to estimate the eigenfunctions or revert to numerical simulations of the process to estimate the FPTD.

Finally, the result of Eq. $(\ref{FPTDtailchi})$ shows that the tail of the FPTD for power-law correlated stationary Markovian processes is different from what is expected for other non-Markovian processes with anomalous diffusion, such as the Fractional Brownian motion. This result might be used to test the Markovian property of a generic stochastic process.


\end{document}